\documentclass[conference]{IEEEtran}
\IEEEoverridecommandlockouts
\usepackage{balance}
\usepackage{comment}
\usepackage{svg}

\usepackage{graphicx}
\usepackage{cite}
\usepackage{amsmath,amssymb,amsfonts}
\usepackage{algorithmicx}
\usepackage{textcomp}
\usepackage{xcolor}

\usepackage{amsthm}

\usepackage[caption=false]{subfig}

\usepackage{authblk}

\usepackage{algorithm,algpseudocode}

\usepackage{multicol}

\usepackage{mathtools}

\usepackage{cuted}

\usepackage{color,soul}
\def\BibTeX{{\rm B\kern-.05em{\sc i\kern-.025em b}\kern-.08em
    T\kern-.1667em\lower.7ex\hbox{E}\kern-.125emX}}

\makeatletter
\newcommand{\linebreakand}{%
\end{@IEEEauthorhalign}
\hfill\mbox{}\par
\mbox{}\hfill\begin{@IEEEauthorhalign}
}

\makeatother

\begin{document}

\title{
Adaptive Dual-Windowing Strategies for Multi-Target Detection in OFDM ISAC

\thanks{This work has been supported by the Smart Networks and Services Joint Undertaking (SNS JU) project 6G-DISAC under the EU’s Horizon Europe research and innovation program under Grant Agreement no. 101139130. 

}
}

\author[1]{Ali Al Khansa}
\author[1, 2]{Youssef Bahannis}

\affil[1]{Orange Labs, Rennes, France}
\affil[2]{INSA, Rennes, France}

\affil[ ]{Emails: \{ali.alkhansa;youssef.bahannis\}@orange.com}

\maketitle

\IEEEpubidadjcol

\begin{abstract}

In Orthogonal Frequency Division Multiplexing (OFDM) Integrated Sensing and Communication (ISAC) systems, a key challenge is balancing sidelobe attenuation and resolution for multi-target detection scenarios. While windowing functions are typically used to manage this trade-off, state-of-the-art methods rely on a single, fixed window followed by a predefined detection strategy (e.g., Binary Successive Target Cancellation (BSTC) (low complexity) or Coherent Successive Target Cancellation (CSTC) (high performance)). This paper proposes a novel dual-window periodogram-based algorithm that leverages two complementary windows: one optimized for resolution and the other for sidelobe suppression. Then, a low-complexity detection algorithm (e.g., BSTC) is applied to both, and a decision mechanism compares the outputs. When results align, the resolution-optimized estimates are directly used; otherwise, high performance algorithm (e.g., CSTC) is triggered to resolve ambiguities. This adaptive approach dynamically balances the detection performance and the complexity, addressing limitations in existing fixed strategies. The Numerical results confirm that the proposed method achieves high performance while reducing the complexity, especially at a high Signal to Noise Ratio (SNR).

\end{abstract}

\vspace{0.5cm}

\section{Introduction}

\IEEEPARstart{I}{\lowercase{n}}tegrated Sensing and Communication (ISAC) has emerged as a key enabler in modern wireless systems, where the dual functions of radar sensing and communication are integrated into a single platform. The advent of ISAC is driven by the growing demands for spectrum efficiency, resource optimization, and the convergence of technologies in 5G and beyond. By enabling seamless coexistence and mutual enhancement of sensing and communication capabilities, ISAC paves the way for innovations in autonomous vehicles \cite{10944644, zhu2024irs}, smart cities, and ubiquitous connectivity, making it a cornerstone for the development of next-generation wireless systems \cite{strinati2024distributed}.

Orthogonal Frequency Division Multiplexing (OFDM) stands out as the modulation scheme of 5G networks and is highly anticipated to play a central role in 6G. Its inherent orthogonality, resilience to multipath fading, and ability to flexibly allocate resources make OFDM a prime candidate for ISAC applications. Furthermore, OFDM's compatibility with Massive MIMO and millimeter-wave technologies reinforces its potential in shaping the future ISAC paradigms \cite{smeenk2024optimizing}. Reflecting these properties, the 3rd Generation Partnership Project (3GPP) \cite{3GPPRAN1} has agreed on an initial 6G baseline that retains Cyclic Prefix (CP)-OFDM for the downlink and adopts Discrete Fourier Transform-spread  (DFT-s)-OFDM for the uplink, marking an early milestone that cements OFDM’s central role in the 6G physical layer \cite{mcns_6g_agreement_2025}.

Among the algorithms designed for spectral estimation in radar, the periodogram method is particularly noteworthy due to its simplicity and effectiveness \cite{mohr2024measurement}. Compared to advanced techniques like MUSIC\footnote{Multiple Signal Classification} \cite{6140075} or ESPRIT\footnote{Estimation of Signal Parameters via Rotational Invariance Techniques} \cite{stoica1997introduction}, periodogram-based methods offer a compelling balance of computational efficiency and estimation performance. Broader options include Bayesian estimators (e.g., \cite{jafri2024bayesian}), however, their performance often comes with higher computational load, model-order/tuning sensitivity, and implementation overhead. Hence, our focus will be on the periodogram-based strategies as a low-complexity, deployment-friendly baseline.

One of the fundamental challenges in radar systems is balancing sidelobe attenuation and resolution. High sidelobe attenuation minimizes interference and enhances weak target detection, while fine resolution improves the system’s ability to distinguish closely spaced targets. This trade-off becomes critical in multi-target scenarios, where resource allocation strategies and sophisticated radar algorithms must carefully navigate these conflicting objectives \cite{li2024low}.

A window function in a periodogram is a weighting function used to balance the performance of the estimation \cite{yu2024rapid}. Specifically, the use of windowing functions within the periodogram framework provides an additional degree of flexibility, enabling fine-tuning of the trade-off between resolution and sidelobe attenuation \cite{braun2014ofdm}. By applying appropriately designed window functions, the performance of OFDM radar can be tailored to specific application requirements, further enhancing its practical utility in challenging sensing environments.

For multi-target detection in OFDM radar, successive cancellation algorithms 
are commonly employed. The Binary Successive Target Cancellation (BSTC) algorithm is computationally simpler, whereas the Coherent Successive Target Cancellation (CSTC) algorithm offers higher performance but comes with significantly greater complexity \cite{braun2014ofdm}. Both algorithms are well-suited for distinct scenarios, but their application is often predetermined, which may not fully leverage their respective advantages in dynamic settings.

Recent years have witnessed significant advances in the integration of sensing and communication functionalities. For instance, \cite{braun2014ofdm} pioneered the use of OFDM waveforms for joint radar and communication, demonstrating the feasibility of periodogram-based spectral estimation for automotive radar applications. Building on this, several works (e.g., \cite{wymeersch2022radio,keskin2021peak,10849719}) explored the trade-offs between resolution and sidelobe levels in OFDM radar. The work of \cite{wymeersch2022radio} shows how different power allocations can lead to significant variation in the sensing performance. Similarly,  the work of \cite{keskin2021peak} introduces a waveform design that allocates power according to Peak Side-lobe Level (PSL) constraints. Recently in \cite{10849719}, a dynamic resource allocation was proposed to achieve different communication and sensing constraints. The common between these approaches is the aim to achieve a balance in the trade-off between resolution and sidelobe attenuation, by using power allocation that acts as a windowing function to handle the trade-off.

Despite these advances, existing approaches typically rely on static window functions and fixed target-detection strategies, which impose a rigid trade-off between sidelobe attenuation and resolution. A further limitation is their capacity cost: power allocation tailored for sensing often departs from the communication-optimal water-filling allocation \cite{cover1999elements} and therefore operates under an explicit data-rate budget, trading communication throughput for sensing constraints.

This work addresses these critical gaps with a capacity-neutral design. We propose a methodology that combines two complementary window functions, one optimized for resolution and the other for sidelobe attenuation, while leaving the transmit spectrum unchanged (i.e., preserving water-filling and thus the communication rate for a given channel state). Furthermore, we introduce a dynamic detection strategy where low-complexity detection algorithm (e.g., the BSTC) is employed by default, with a seamless transition to a more sophisticated detection algorithm (e.g., CSTC) only when necessary. Specifically, we perform a matching process between the estimation results of the two windows used, and then, according to the decision of the matching process, we decide whether there is a need to perform a more complex detection algorithm or not. This adaptive framework not only eliminates the traditional trade-off but also optimally balances performance and complexity, without consuming a capacity budget, paving the way for more efficient and robust OFDM radar systems in ISAC applications.

The remainder of this paper is organized as follows: Section II provides an overview of OFDM radar and periodogram-based spectral estimation. Section III highlights the effect of windowing functions, focusing on the trade-off between sidelobe attenuation and resolution, and discusses the BSTC and the CSTC algorithms. Section IV introduces our proposed methodology, which leverages dual-windowing and adaptive successive cancellation strategies to overcome the limitations of existing approaches. Numerical results validating the effectiveness of the proposed method are presented in Section V. Finally, Section VI concludes the paper.

\section{OFDM ISAC with Periodogram-Based Estimation Algorithm}

Before presenting our proposal, we start by presenting the OFDM radar fundamentals, followed by the periodogram-based estimation, so that we show how windowing functions can be applied. 

\subsection{OFDM Radar Fundamentals}
In an OFDM radar system, the received signal $r(t)$ is expressed as a superposition of reflections from $H$ targets, each with unique characteristics such as attenuation, time delay, and Doppler shift. The received signal can be modeled as \cite{braun2014ofdm}:
\begin{equation}
    r(t) = \sum_{h=0}^{H-1} b_h s(t-\tau_h) e^{j 2 \pi f_{D,h} t} e^{j \overline{\phi}_h} + \overline{z}(t),
\end{equation}
where $b_h$ represents the attenuation factor, $\tau_h$ is the round-trip delay, $f_{D,h}$ is the Doppler frequency shift, $\overline{\phi}_h$ is a random phase offset, and $\overline{z}(t)$ denotes Additive White Gaussian Noise (AWGN). The transmitted signal $s(t)$, as defined in the OFDM scheme, comprises subcarriers that are orthogonal in time and frequency, ensuring efficient spectral utilization and minimizing interference.


To apply the previous equation to OFDM signals specifically, let us introduce a new notation.
A transmitted OFDM frame is represented by a matrix:
\begin{equation}
\label{F_Tx}
    \textbf{\textit{F}}_{T_x} = \begin{pmatrix}
c_{0,0} & \cdots & c_{0,M-1} \\
\vdots & \ddots & \vdots \\
c_{N-1,0} & \cdots & c_{N-1,M-1}
\end{pmatrix},
\end{equation}
where symbols $c_{k,l}$, $k\in\{0,\dots,N-1\}$ and $l\in\{0,\dots,M-1\}$ are chosen from a modulation alphabet (e.g., QPSK, QAM, etc.), $N$ is the number of subcarriers and $M$ is the number of OFDM symbols. Each row of $\textbf{\textit{F}}_{T_x}$ represents a subcarrier and each column represents an OFDM symbol of the transmitted frame. The received frame matrix $\textbf{\textit{F}}_{R_x}$, derived through analog-to-digital conversion and demodulation of the received signal, captures the effects of the propagation channel. For a single target, i.e., $H=1$, $\textbf{\textit{F}}_{R_x}$ can be written as:
\small \begin{equation}
\label{F_Rx}
(\textbf{\textit{F}}_{R_x})_{k,l} = b_0(\textbf{\textit{F}}_{T_x})_{k,l} e^{j 2 \pi T_O f_{D,0}l} e^{-j 2 \pi \tau_0(k\Delta f + f_0)} e^{j \overline{\phi}_0} + (\overline{\textbf{\textit{Z}}})_{k,l},
\end{equation} \normalsize 
where $\overline{\textbf{\textit{Z}}} \in \mathbb{C}^{N \times M}$ is the matrix representation of the AWGN, $f_0$ is the initial frequency of the $N$ subcarriers (i.e., the frequencies go from $f_0$ through $f_{N-1}$), $T_O$ is the OFDM symbol duration (including the duration of the cyclic prefix), and $\Delta f$ is the subcarrier spacing. As $f_0$ and $\overline{\phi}_0$ are constant, define $\phi_h=\overline{\phi}_h-2\pi f_0 \tau_h$. To isolate the effects of target reflections, element-wise division of the received matrix by the transmitted matrix produces the normalized frame matrix \textbf{\textit{F}}, expressed as: 
\begin{equation}
\label{F_0}
    (\textbf{\textit{F}})_{k,l} = b_0 e^{j 2 \pi l T_O f_{D,0}} e^{-j 2 \pi k \tau_0 \Delta f} e^{j \phi_0} + (\textbf{\textit{Z}})_{k,l},
\end{equation}
where $(\textbf{\textit{Z}})_{k,l}= (\overline{\textbf{\textit{Z}}})_{k,l} / \textbf{\textit{F}}_{T_x}$ represents the normalized AWGN. This formulation transforms the radar estimation problem into a spectral estimation problem, where the parameters of interest, i.e., the delay $\tau$ and the Doppler frequency $f_D$, correspond to the target's distance and relative velocity, respectively.

Now, this result can be generalized to $H>1$ targets, since the operations used to calculate $\textbf{\textit{F}}_{R_x}$ from $r(t)$ are linear with respect to their input signal:
\begin{equation}
\label{F_general}
    (\textbf{\textit{F}})_{k,l} = \sum_{h=0}^{H-1} b_h e^{j 2 \pi l T_O f_{D,h}} e^{-j 2 \pi k \tau_h \Delta f} e^{j \phi_h} + (\textbf{\textit{Z}})_{k,l}.
\end{equation}

\subsection{Periodogram-Based Estimation}

\begin{figure}[t]
\centering
\includegraphics[scale=0.26]{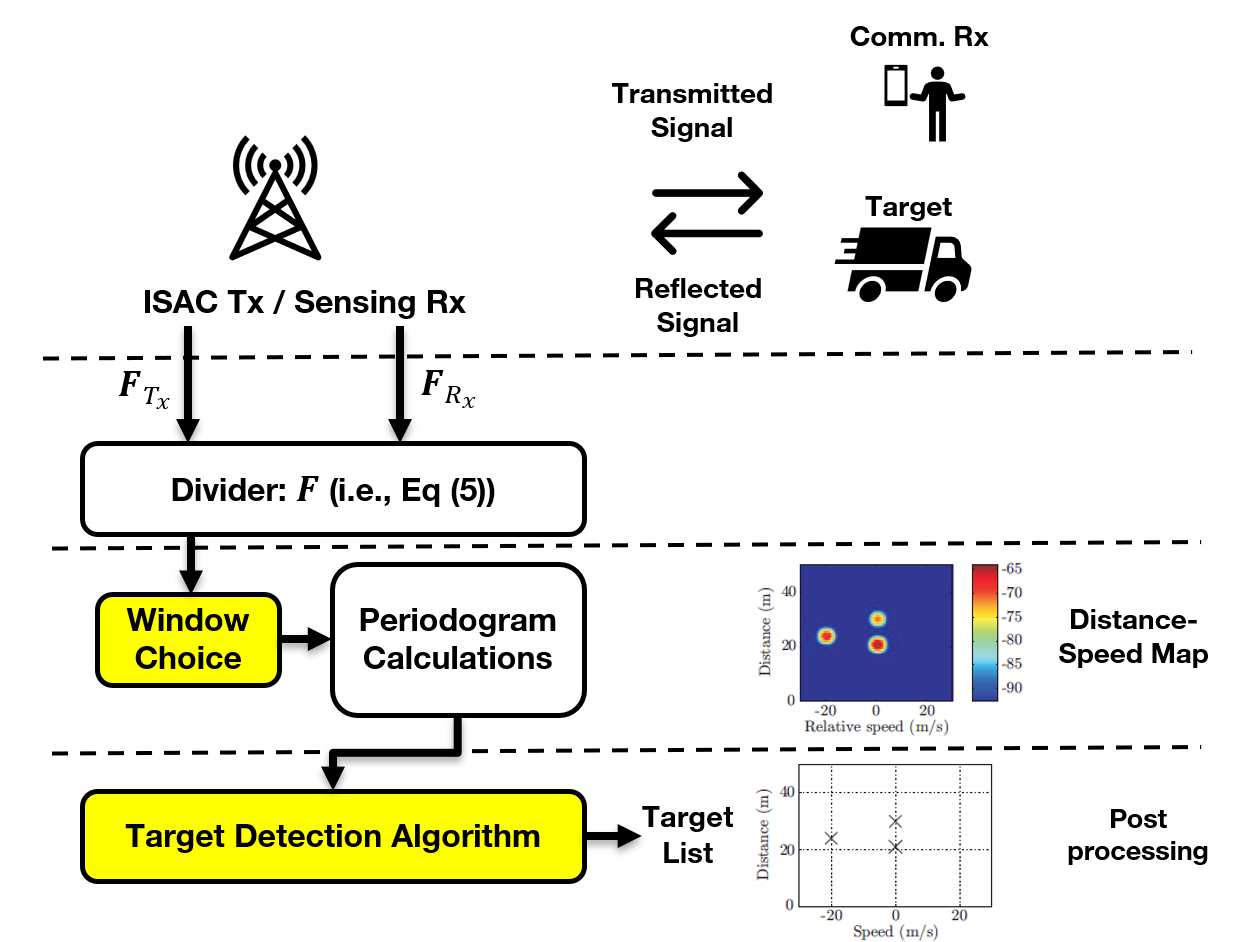}
\caption{Block diagram of a periodogram-based OFDM ISAC system. The contributions of this paper relate to the highlighted yellow parts: the windowing and the target detection algorithm.}
\label{fig:3.6}
\end{figure}

To estimate the target parameters from the matrix $\textbf{\textit{F}}$, the periodogram is used as a spectral estimation tool. The two-dimensional periodogram for the received data is defined as (Eq. (3.30) \cite{braun2014ofdm}):
\begin{equation}
\label{Per_F}
    \text{Per}_\textbf{\textit{F}}(n, m) = \frac{1}{NM} \left| \sum_{k=0}^{N_{\text{Per}}-1} \sum_{l=0}^{M_{\text{Per}}-1} (\textbf{\textit{F}})_{k,l} e^{-j 2 \pi \frac{l m}{M_{\text{Per}}}} e^{j 2 \pi \frac{k n}{N_{\text{Per}}}} \right|^2,
\end{equation}
where $N_{\text{Per}}$ and $M_{\text{Per}}$ represent the size of the Inverse Fast Fourier Transforms (IFFT) and the Fast Fourier Transforms (FFT). These transforms are applied to the rows and columns of the matrix \textbf{\textit{F}}, isolating the sinusoidal components corresponding to the targets' Doppler shifts and delays. Peaks in the periodogram represent the estimated parameters, which can then be translated into target distances and velocities. The periodogram-based algorithm is computationally efficient, relying on FFT operations and simple peak detection, which makes it particularly suitable for real-time radar applications. 


%

Fig. \ref{fig:3.6} illustrates a block diagram of a periodogram-based OFDM ISAC system. The diagram begins with a monostatic ISAC transceiver that transmits a signal to a communication user while simultaneously receiving reflected signals from a target. The signal processing chain starts by dividing the received signal by the transmitted signal, as described in Eq. (\ref{F_general}). Following this, the periodogram is calculated to generate the range–Doppler map. A target detection algorithm is then applied to extract the target list from this map. The two components highlighted in yellow, the window choice and the target detection algorithm, indicate the parts of the system most related to the contributions of this paper and will be discussed in more detail in the following section.



\section{Windowing Effect and Trade-off Analysis in Multi-Target Scenarios}

\subsection{Windowing for Resolution-Sidelobe Trade-off}
As previously explained, a fundamental challenge in radar signal processing is the trade-off between range resolution and sidelobe attenuation. High range resolution allows the system to distinguish closely spaced targets, while strong sidelobe attenuation ensures that weaker targets are not masked by sidelobes from stronger reflections. 

Window functions have been widely adopted as a solution to fine-tune the balance between resolution and sidelobe attenuation. A window function modulates the input signal, altering its spectral characteristics. For instance, using a rectangular window provides a narrow main lobe (optimized for resolution and close targets detection) which is usually considered as a benchmark (rectangular window is equivalent to no window at all \cite{braun2014ofdm}). However, a rectangular window suffers from high sidelobe levels, which can obscure weak targets in the presence of strong ones. Conversely, applying a window function such as Hamming or Blackman-Harris reduces sidelobes but broadens the main lobe, degrading resolution. This inherent trade-off complicates the design of algorithms that must perform well across diverse sensing scenarios. Commonly used window functions and their designs can be checked in \cite{braun2014ofdm}, and a short summary of these windows is listed below:
\begin{itemize}  
    \item  Rectangular Window: Offers a good resolution with a narrow main lobe but has poor sidelobe attenuation (13 dB).
    \item  Hamming Window: Improves sidelobe attenuation (42 dB) at the cost of doubling the main lobe width.
    \item Dolph-Chebyshev Window: Features user-defined sidelobe levels and a main lobe width that adapts to the specified attenuation.
\end{itemize}

Now, to implement a window function in periodogram, a two-dimensional window matrix $\textbf{\textit{W}}$ is defined, which is multiplied element-wise with the matrix $\textbf{\textit{F}}$. Accordingly, equation (\ref{Per_F}) can be rewritten as:
\begin{align}
    \label{Per_F_Window}
    & \text{Per}_\textbf{\textit{F}}(n, m) =  \\ \nonumber
    & \frac{1}{NM} \left| \sum_{k=0}^{N_{\text{Per}}-1} \sum_{l=0}^{M_{\text{Per}}-1} (\textbf{\textit{F}})_{k,l}(\textbf{\textit{W}})_{k,l} e^{-j 2 \pi \frac{l m}{M_{\text{Per}}}} e^{j 2 \pi \frac{k n}{N_{\text{Per}}}} \right|^2.
\end{align}

\begin{figure}
\centering
\includegraphics[scale=0.48]{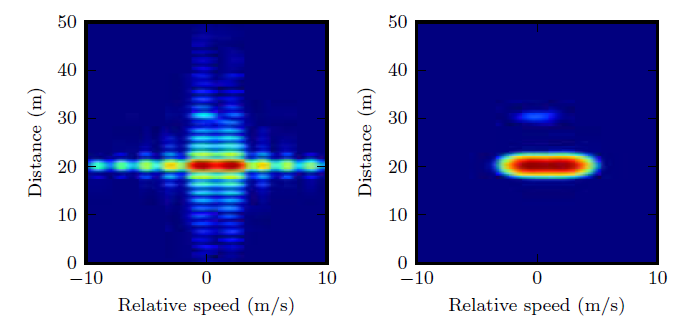}
            \caption{Periodogram of three targets using two different windows: (a) rectangular window optimized for resolution: the two targets at 20m are easily distinguished, (b) Dolph-Chebyshev window optimized for sidelobe attenuation: the target at 30m now clearly stands out \cite{braun2014ofdm}.}
\label{fig:3.10'}
\end{figure}

Fig. \ref{fig:3.10'}  illustrates the effect of windowing functions on the periodogram, highlighting the trade-off between resolution and sidelobe attenuation and its impact on target detection. It presents two periodograms for identical scenarios, each using a different window function. In the first case with a rectangular window, the main lobe is narrow, enabling the distinction of closely spaced targets (higher resolution). In the second case, a window with strong sidelobe suppression is used, which allows weaker targets to be distinguished from the sidelobes of stronger ones, but at the cost of reduced resolution. The contribution of this paper lies in leveraging a dual-window approach that simultaneously considers both resolution and sidelobe attenuation, thereby balancing detection performance across diverse conditions.

\subsection{Multi-Target Detection Algorithms}
A target detection algorithm is applied after the periodogram calculations and is responsible for extracting the list of detected targets along with their characteristics, such as distance and speed, from the range–Doppler map. In multi-target scenarios, successive cancellation algorithms (e.g., the BSTC and the CSTC) are commonly employed for target detection. These methods detect one target per iteration and cancel its effect to facilitate the detection of the remaining targets. The BSTC assumes high sidelobe attenuation (e.g., using Hamming window) and is computationally simpler, relying on the suppression of sidelobes to iteratively identify and remove the strongest targets. In contrast, the CSTC is more sophisticated and takes into account the phase coherence of the radar signal when performing the successive cancellation. This helps to accurately cancel the multiple targets and their sidelobes (with no need of a window), but comes with a cost of added complexity. While both algorithms are effective in different contexts, their application is typically fixed, which can limit adaptability in dynamic conditions. These two methods are not presented in detail in this paper due to brevity, and readers are referred to \cite{braun2014ofdm} for more information. 


\section{The Dual-Window Proposal}
\subsection{The Proposed Strategy}



To overcome the limitations of using a single window and static cancellation algorithm, we propose a novel dual-windowing approach coupled with an adaptive cancellation strategy. 
The dual-windowing strategy uses two complementary window functions, instead of relying on a single window. The first window is optimized for resolution (e.g., a rectangular window) and the second window is optimized for sidelobe-attenuation reduction (e.g., Dolph-Chebyshev). This approach enables us to exploit the strengths of each window while minimizing its respective drawbacks. Upon using two windows, the dynamic selection algorithm for the multi-target detection begins by using a simple, low-complexity cancellation algorithm(s) for each window (e.g., use the BSTC algorithm for both windows). Then, based on the estimated results from the two windows and two cancellation algorithms, we dynamically select the appropriate result. Here, we consider two cases. The first case is to have what we call \textit{convergence} in the results, or consistent results, meaning that each estimation of the two windows is giving a set of targets with range and Doppler values that align with the second window. In this case, our dynamic proposal selects the estimation results of the window optimized for resolution. Now, in case we have ambiguity between the results (results are not consistent), or what we call \textit{divergence} between the estimation of the two windows, then, in this case, we call for a more complex cancellation algorithm (e.g., the CSTC algorithm) to try solving the ambiguity issue we are facing. 

This strategy aims to always take the best resolution, when possible (via using the two windows), and to only perform the more complex algorithm (e.g., the CSTC) when we encounter ambiguity. Note that \textit{ambiguity} here can be detected by comparing the set of estimates we have from using the two windows. If the number of targets and their specifications are matched, we have no ambiguity, and thus, we can use the window optimized for resolution. If this is not the case, i.e., if we have ambiguity (e.g., the number of targets is not the same, the location of targets and/or Dopplers are not close to each other), we need to perform the CSTC algorithm.

\begin{figure}
\centering
\includegraphics[scale=0.29]{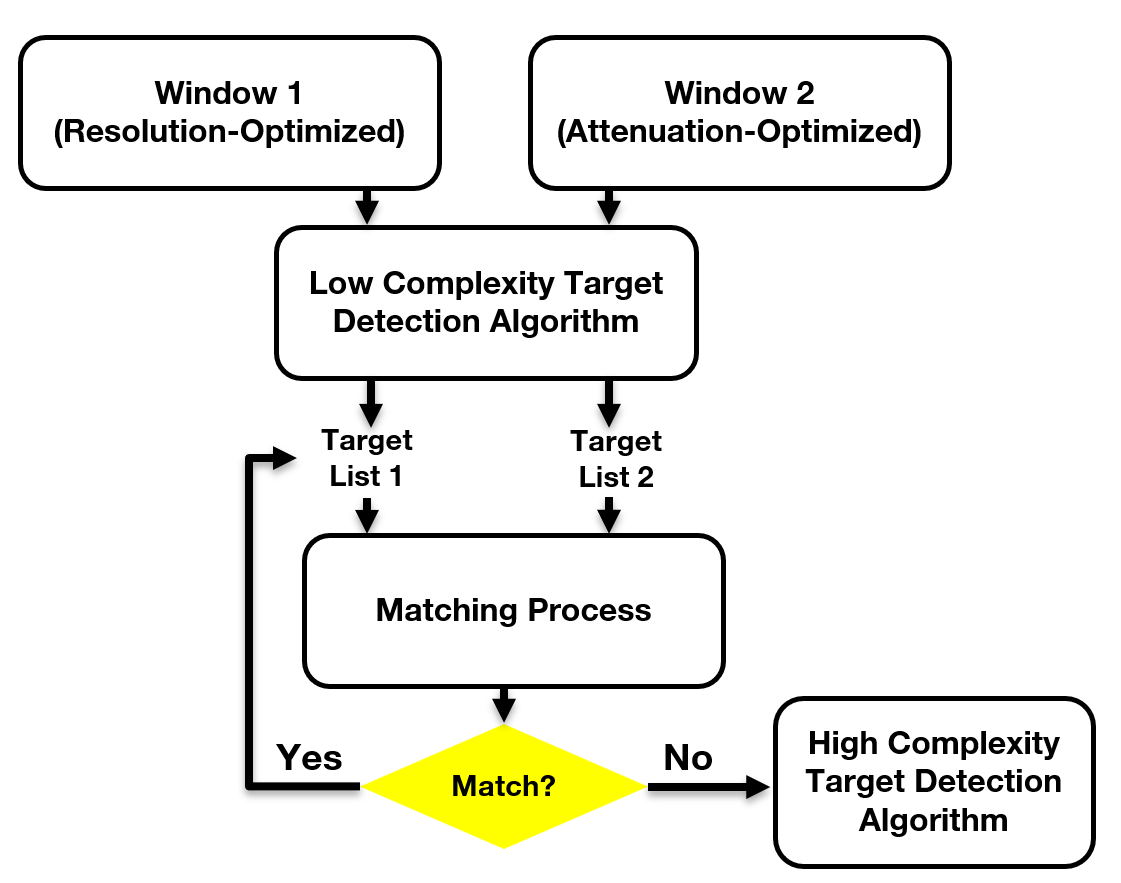}
\caption{Block diagram of the proposed dual-window and adaptive detection framework.}
\label{fig:3.6'}
\end{figure}

Following the logic of the proposed method, we ensure that we start by using a simple algorithm (e.g., the BSTC) and we detect ambiguities (by the matching process). If no ambiguity is detected, a high resolution result is used (window trade-off handled). And on the contrary, if ambiguity is detected, we try suppressing it via a more complex cancellation algorithm and this is the only case where we use such complex algorithms (complexity trade-off handled). 
Fig. \ref{fig:3.6'} illustrates a block diagram of the proposed dual-window and adaptive detection framework. Periodogram calculations and low-complexity detection algorithms are applied using two different window types. The resulting target lists are compared through a matching process. If the results match, the high-resolution output is selected; otherwise, a high-complexity detection algorithm is triggered to resolve ambiguities.



\subsection{The Matching Process}
We see that in the previous steps, we talked about the matching of the results of the estimation. Let us explain in more detail how exactly this process can be done. As mentioned above, if $N_\text{res} \neq N_\text{sidelobe}$ (number of targets of using resolution-optimized window and sidelobe attenuation-window), we directly deduce that we do not have a match between the two estimates, and thus, we have ambiguity, and we need to perform a complex cancellation algorithm. However, if $N_\text{res} = N_\text{sidelobe}$, we cannot directly deduce that we have matching results, and we need to check the ranges and the Doppler matching. We explain this case below.

To confirm that we have a matching pair of range and Doppler, we need to pair targets from the two estimates of the two different windows, say estimate 1 and estimate 2. Using the optimal \textit{Nearest Neighbor Matching} strategy, or the practical \textit{Hungarian Algorithm} \cite{kuhn1955hungarian} (suitable for a large number of targets), we will reach $N_\text{pairs} =  N_\text{res} = N_\text{sidelobe}$ of estimates, each containing a pair of range and Dopplers: (($\text{range}_{\text{res}_i},\text{Doppler}_{\text{res}_i}$), ($\text{range}_{\text{sidelobe}_i},\text{Doppler}_{\text{sidelobe}_i}$)) for all $i \in \{1,\dots, N_\text{pairs}\}$. Finally, using a distance-like metric, e.g., the squared Euclidean distance, we can compute the maximum difference between the different pairs as:
\begin{align}
    d = \max_{i \in \{1,\dots, N_\text{pairs}\}} &  \left( \text{range}_{\text{res}_i} - \text{range}_{\text{sidelobe}_i} \right)^2 +  \nonumber \\   
    & \left( \text{Doppler}_{\text{res}_i} - \text{Doppler}_{\text{sidelobe}_i} \right)^2
\end{align}
If $d\leq \epsilon$, we declare a matching result, otherwise, we declare a non-matching result. Note that $\epsilon$ is a system parameter that depends on the application. Algo. \ref{algo} represents the pseudocode for the proposed method. Note that in step 3, we assume that a matching algorithm is being used to determine $d$ as explained in the previous example.

\begin{algorithm}[ht]


\textbf{Step 1: Compute periodogram using:}\\
\hspace{1cm} a. Resolution-optimized window $\rightarrow \text{Per}_\text{res}$\\
\hspace{1cm} b. Sidelobe-attenuation window $\rightarrow \text{Per}_\text{sidelobe}$\\

\textbf{Step 2: Apply simple cancellation algorithm (e.g., the BSTC) for both windows:}\\
\hspace{1cm} a. $N_{\text{res}}$ estimates of ranges and Dopplers $\left(\text{range}_{\text{res}_i},\text{Doppler}_{\text{res}_i}\right)$, for $i \in \{1,\dots, N_\text{res}\}$  \\
\hspace{1cm} b. $N_{\text{sidelobe}}$ estimates of ranges and Dopplers $\left(\text{range}_{\text{sidelobe}_i},\text{Doppler}_{\text{sidelobe}_i}\right)$, for $i \in \{1,\dots, N_\text{sidelobe}\}$  \\

\textbf{Step 3: Compare target counts:}\\
\hspace{1cm} \textbf{if} ($N_\text{res} == N_\text{sidelobe}$ and $d \leq \epsilon$) \textbf{then}\\
\hspace{2cm} We declare a matching result and thus we use the estimates from the resolution-optimized window\\
\hspace{1cm} \textbf{else}\\
\hspace{2cm} We declare a non-matching result and thus we perform a more complex cancellation algorithm (e.g., the CSTC). \\
\caption{Proposed Multi-Target Detection Algorithm}
\label{algo}
\end{algorithm}






\section{Numerical Results}
\subsection*{Setup and Results}
In this section, we validate our proposed algorithm via numerical simulations using an OFDM model with $N=64$, $M=256$, $N_{\text{Per}}=4N$, and $M_{\text{Per}}=4M$. A 16-QAM modulation is considered with $T_O=4 \mu$s, $\Delta f = 312.5$ kHz, and $f_0 =5.5$ GHz. We investigate the performance of our proposal, compared to three benchmark algorithms: the BSTC with a rectangular window optimized for resolution, the BSTC with a 80-dB Dolph–Chebyshev window optimized for sidelobe attenuation, and the CSTC \cite{braun2014ofdm}. Note that any other pair of windows can be used, with the condition that the first one is optimized for resolution and the second one is optimized for sidelobe attenuation. A 5000 Monte-Carlo simulation of a three-target scenario is considered, where the RCS of these targets are fixed to $10 \text{ m}^2$ and the distance of these targets are generated randomly in the region of [10 m, 80 m] and the speeds of [-100 \text{m/s}, 100 \text{m/s}] separated by a minimum of 10 m and 10\text{ m/s} from each other. Accordingly, we fix $\epsilon$ to 10.

\begin{figure}
    \centering
    \includegraphics[scale=0.58]{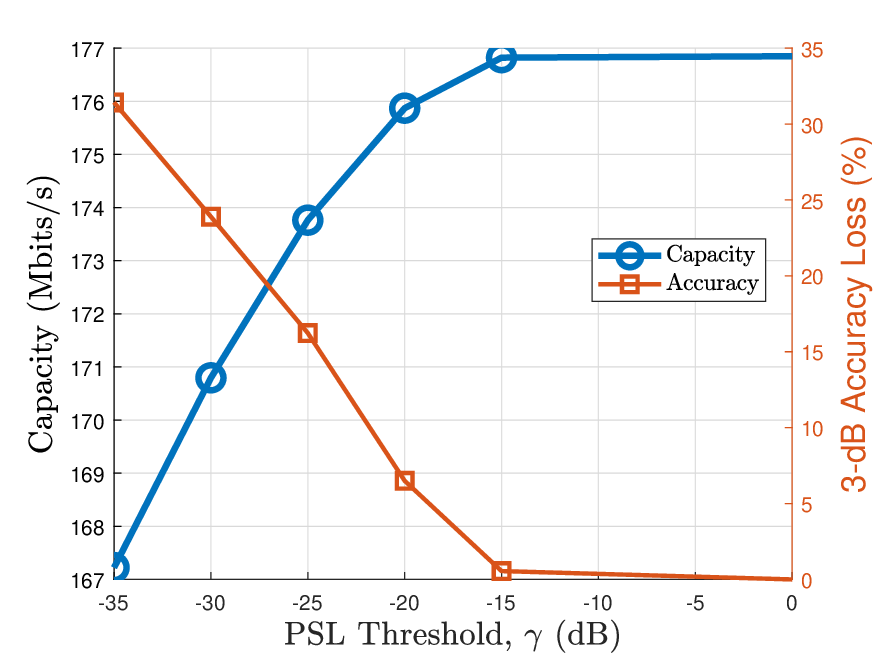}
    \caption{Trade-offs between capacity and accuracy in OFDM ISAC Systems under different PSL constraints (128 subcarriers, 1MHz bandwidth) \cite{10849719}}
    \label{CSCNFig2}
    \end{figure}

Before presenting our results, we briefly revisit the capacity-sensing trade-off studied in \cite{keskin2021peak,10849719}, where power allocation is optimized subject to a PSL constraint. In \cite{keskin2021peak}, the power allocation departs from optimal water-filling to satisfy a PSL threshold $\gamma$, thereby maximizing capacity under that constraint. In \cite{10849719}, we extend this with a scenario-dependent (dynamic) allocation that selects a point along the same trade-off. Figure \ref{CSCNFig2} illustrates this tradeoff: as $\gamma$ becomes more stringent, capacity decreases (blue curve, circles) and the 3-dB accuracy loss increases. This baseline motivates our dual windowing strategy performed at the receiver side. In contrast to power allocation methods, our dual-windowing targets the resolution–sidelobe compromise directly at the receiver, using two complementary windows and adaptive selection. Crucially, because the power allocation can always be fixed via water-filling, our method is capacity-neutral: it operates at the maximum capacity for the given channel (blue curve, circles) while targeting the resolution-ambiguity tradeoff via the dual windowing method (red curve, squares). This framing clarifies the role of our approach: it addresses the sensing trade-off without consuming a data-rate budget.

\begin{figure}
    \centering
    \subfloat[Detection Performance.]{%
        \includegraphics[width=0.48\textwidth]{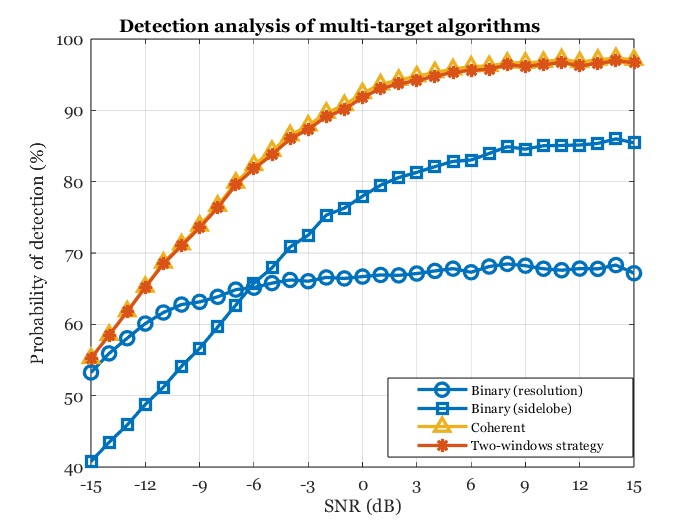}
        \label{fig:NRfig_0a}
    }
    \hfill
    \subfloat[Complexity Performance.]{%
        \includegraphics[width=0.48\textwidth]{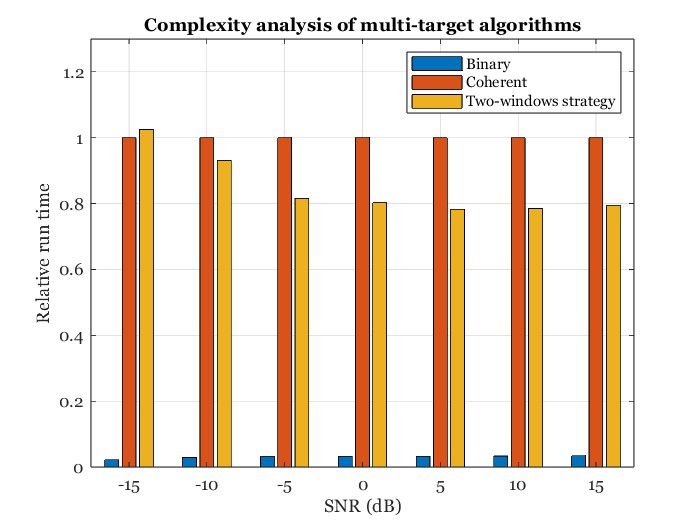}
        \label{fig:NRfig_0b}
    }
    \caption{Comparative analysis of performance and complexity of multi-target detection strategies for different SNR values.}
    \label{fig:combined_figure}
\end{figure}

Next, we present the performance of our proposal. In Fig. \ref{fig:combined_figure}, we present the detection probability (i.e., the probability of estimating the range and the Doppler of a target correctly, with a tolerance less than 5m and 5\text{m/s} respectively) and the complexity of the proposed algorithm compared to three benchmark methods. Although full details of the BSTC and CSTC algorithms are beyond the scope of this work, we briefly highlight key considerations used in our simulations.

Regarding the detection performance (Fig. \ref{fig:NRfig_0a}), an important aspect is the stopping criterion for detecting targets. This typically involves deciding whether a peak in the periodogram corresponds to a real target or noise. Various thresholding methods exist, including fixed thresholds based on noise levels and adaptive approaches such as Constant False Alarm Rate (CFAR). Following the approach in \cite{braun2014ofdm}, where an ideal CFAR is assumed, we simulate under the assumption that the number of targets is known. This allows us to isolate the algorithm's performance in resolving ambiguities caused by limited resolution and sidelobe interference.

For the complexity analysis (Fig. \ref{fig:NRfig_0b}), we report the runtime of each algorithm normalized by the CSTC runtime. For the proposed method, we sum the runtime of the two BSTC runs and include the CSTC runtime only in cases where divergence is detected in the matching process. The cost of the matching process itself is considered negligible, as the number of targets is small (three).

In Fig. \ref{fig:NRfig_0a}, the CSTC achieves the highest detection probability reaching above 95\% at a high Signal to Noise Ratio (SNR), and our proposed algorithm closely matches its performance, while clearly outperforming both BSTC variants. All four methods saturate at high SNR due to multi-target ambiguities; in a single-target case (not shown), performance reaches 100\%. This highlights the importance of considering the ambiguities from multiple targets. Fig. \ref{fig:NRfig_0b} shows that BSTC has significantly lower complexity than CSTC, while the proposed method exhibits SNR-dependent complexity: it is higher at low SNR and decreases as SNR increases (reaching around 80\% of CSTC complexity at high SNR).

This behavior is expected. At low SNR, noise affects the convergence of the matching process, causing the proposed method to frequently fall back to CSTC, thus increasing the total complexity. As SNR improves, BSTC becomes more reliable, reducing the need for CSTC and lowering overall complexity. Therefore, the proposed strategy offers CSTC-level detection performance but only incurs CSTC-level complexity when needed, making it more efficient at moderate to high SNR.

\subsection*{Limitations and Future Works}

Despite its promising performance, our approach has limits. First, complexity gains are SNR-dependent: at low SNR the matching stage often triggers the high-complexity fallback (e.g., CSTC), eroding runtime benefits; only at higher SNR do we reliably realize low-complexity operation. Second, our experiments used small target counts; as scene density increases, assignment/matching scales non-trivially and can dominate runtime unless optimized.

To mitigate these limitations, we are developing an extension of the proposed strategy that preserves the dual-window principle while replacing explicit target matching with a learning model. Specifically, we feed the two windowed range–Doppler periodogram maps to a Machine Learning (ML) model (e.g., Convolutional Neural Network (CNN)) that directly outputs target estimates, thereby avoiding the matching overhead that grows with the number of targets. As the CNN model is trained offline, we avoid the complexity penalty encountered in the matching process of our proposal. Recent CNN-based estimators for OFDM radar show promise \cite{choi2021multiple,choi2022estimation,jeon2024velocity}, but they typically use a single-window representation, inheriting the same resolution–sidelobe trade-off and degrading as scene density increases. 

In \cite{choi2021multiple,choi2022estimation}, several CNN architectures ranging from 4- to 7-layer models were evaluated with the objective of inferring the number of targets from a periodogram input. The key observations were: (i) a clear performance–complexity trade-off across models, and (ii) declining accuracy as the number of targets increases. A related study in \cite{jeon2024velocity} explored a YOLO-based detector and reported broadly comparable trends.

We note that we can explore more expressive models without strict inference-time complexity penalties as training occurs offline. Our emphasis is therefore on improving estimation accuracy, especially in high-density scenes. Thus, we will extend the single-map formulations by using the proposed dual-window inputs to boost robustness as the target count grows. We also observe that \cite{choi2021multiple,choi2022estimation,jeon2024velocity} primarily address target-count estimation rather than joint range–Doppler localization, which we identify as an additional limitation to overcome. Future work will thus target end-to-end prediction of both target number and range–Doppler states directly from the dual-map inputs. 
 We leave detailed analysis/results for future work (e.g., \cite{Learning}) due to space constraints and because the ML extension lies beyond this paper’s scope.

\section{Conclusions}
This paper introduced a dual-window periodogram-based algorithm for OFDM radar in ISAC systems, addressing the trade-off between resolution and sidelobe attenuation. By combining two complementary window functions with an adaptive selection strategy, the proposed method dynamically balances detection accuracy and computational complexity. Numerical results confirm that our approach achieves the performance of the CSTC algorithm with reduced processing complexity, making it well-suited for practical ISAC applications. 

\bibliography{conf1}

\end{document}